\documentclass[useAMS,usenatbib]{mn2e}
\usepackage{graphicx}
\usepackage{float}
\usepackage{wasysym}
\usepackage{txfonts}
\usepackage{natbib}
\usepackage{color}
\usepackage{colortbl}
\usepackage{multirow}

\title[Young protoplanetary discs]{Simulated Observations of Young Gravitationally Unstable Protoplanetary Discs}

\author[T.~A.~Douglas et al.]
{\parbox{\textwidth}{T. A. Douglas$^{1}$\thanks{E-mail: \texttt{pytd@leeds.ac.uk}},
P. Caselli$^{1}$,
J. D. Ilee$^{2,1}$,
A. C. Boley$^{3}$,
T. W. Hartquist$^{1}$,
R. H. Durisen$^{4}$, and
J. M. C. Rawlings$^{5}$\\
\vspace{0.1cm}\\
{\small{\it$^{1}$School of Physics and Astronomy, University of Leeds, Leeds LS2 9JT, UK,}} \\
{\small{\it$^{2}$School of Physics and Astronomy, University of St Andrews, North Haugh, St Andrews, KY16 9SS, UK,}}\\
{\small{\it$^{3}$Department of Astronomy, University of Florida, 211 Bryant Space Center, PO Box 112055, USA,}}\\
{\small{\it$^{4}$Department of Astronomy, Indiana University, 727 East 3rd Street, Swain West 319, Bloomington, IN 47405, USA,}}\\
{\small{\it$^{5}$Department of Physics \& Astronomy, University College London, London WC1E 6BT, UK}}\\
}}

\begin{document}

\date{Accepted May 2013}

\pagerange{\pageref{firstpage}--\pageref{lastpage}} \pubyear{2002}

\maketitle

\label{firstpage}

\begin{abstract}
The formation and earliest stages of protoplanetary discs remain poorly constrained by observations. ALMA will soon revolutionise this field. Therefore, it is important to provide predictions which will be valuable for the interpretation of future high sensitivity and high angular resolution observations. Here we present simulated ALMA observations based on radiative transfer modelling of a relatively massive (0.39 M$_{\odot}$) self-gravitating disc embedded in a 10\,M$_{\odot}$ dense core, with structure similar to the pre-stellar core L1544. We focus on simple species and conclude that  C$^{17}$O 3$\rightarrow$2, HCO$^+$ 3$\rightarrow$2, OCS 26$\rightarrow$25 and H$_2$CO 4$_{04}\rightarrow$3$_{03}$ lines can be used to probe the disc structure and kinematics at all scales. 
\end{abstract}

\begin{keywords}
Stars: Circumstellar Matter -- infrared; Stars: Formation -- infrared; Planetary Systems: Protoplanetary Discs.
\end{keywords}

\section{Introduction}

Currently few observations constrain the formation and early evolution of protoplanetary discs around solar-type and low-mass protostars, despite the fast growing list of theoretical models on the dynamical evolution of star forming dense cores (e. g. \citealt{Krasnopolsky2011,Machida2011,Braiding2012,Joos2013}). The reason for this is that young protostars are surrounded by thick envelopes and power energetic outflows. Thus, observations of the young discs, predicted to have sizes of about 100\,au and masses as large as 10 percent the original core mass (e. g. \citealt{Joos2012,Hayfield2011,Machida2011}), are challenging. Sensitive interferometers are needed to achieve high angular resolution and spatially/spectrally disentangle the various disc, envelope and outflow components as well as to filter out the extended emission tracing the envelope material. \smallskip

After the pioneering work of, e. g., \citet{Chandler1995,Brown2000,Looney2000}, recent interferometric observations have discovered compact embedded discs in a sample of Class 0 sources (as defined by \citealt{Andre1999}), finding masses between 0.4 and  $>$1 M$_{\odot}$ \citep{Jorgensen2007,Jorgensen2009,Enoch2011}. A 130\,au disc was discovered in NH$_3$ emission toward a Class 0 source in Perseus, with the Jansky Very Large Array (JVLA) \citep{Choi2007}. \citet{Pineda2012} observed methyl formate (HCOOCH$_3$) with the Atacama Large Millimetre/sub-millimetre Array (ALMA) and found evidence of rotation toward one of the proto-binary Class 0 sources embedded in IRAS 16293-2422. These observations are consistent with an almost edge-on disc. \citet{Persson2012} observed H$_2^{18}$O with ALMA toward the same source and found evidence of relatively large ($\ga$100\,K) excitation temperatures, as well as a HDO/H$_2$O abundance ratio close to that measured on Earth's oceans and Jupiter-family comets \citep{Hartogh2012}. \citet{Zapata2013} used ALMA to observe a disc of size $\sim$50$\,$au in the other Class 0 source in IRAS 16293-2422 and detected inverse P-Cygni profiles in HCN and CH$_3$OH which indicate infall towards the disc. When ALMA is completed, it will be finally possible to spatially resolve these young discs and, for the first time, put stringent constraints on the theoretical models mentioned previously. As fully-operational ALMA is fast approaching, it is important to provide observational predictions based on dynamical models of young protoplanetary discs. \smallskip 

Simulated observations of gravitationally unstable discs have already been performed to study the continuum emission, measure the structure and investigate possible fragmentation \citep{Cossins2010,Ruge2013}. Molecular line emission from massive discs has been simulated by \citet{Krumholz2007}, who assumed local thermodynamic equilibrium (LTE). In this paper, we focus on the self-gravitating discs of \citet{Boley2008}, in which episodic heating induced by spiral shocks is present. This may be a good representation of the earliest phases of protoplanetary discs and an alternative to the young ``static'' discs studied by, e. g. \citet{Visser2009,Visser2011}. As shown by \citet{Ilee2011}, hereafter I2011, the spiral shocks cause desorption of volatiles from dust icy mantles and trigger gas-phase chemical reactions with activation energies that retard the reaction at lower temperatures. These processes produce clear chemical signatures of the disc dynamics. With the use of non-LTE 3D radiative transfer modelling, we have simulated ALMA observations of the disc studied by I2011 and identify the best tracers of the physical structure of self-gravitating discs. The physical, chemical and radiative transfer models are described in Sect.\,\ref{sec:description_model}. Radiative transfer results are in Sect.\,\ref{sec:model_results}, while ALMA simulated observations are in Sect.\,\ref{sec:alma_predictions}. Discussions and conclusions can be found in Sect.\,\ref{sec:discussion}. 

\section{Description of the Model} \label{sec:description_model}

\subsection{Physical structure} \label{subsec:physical_structure}

\begin{figure}
 \includegraphics[width=84mm]{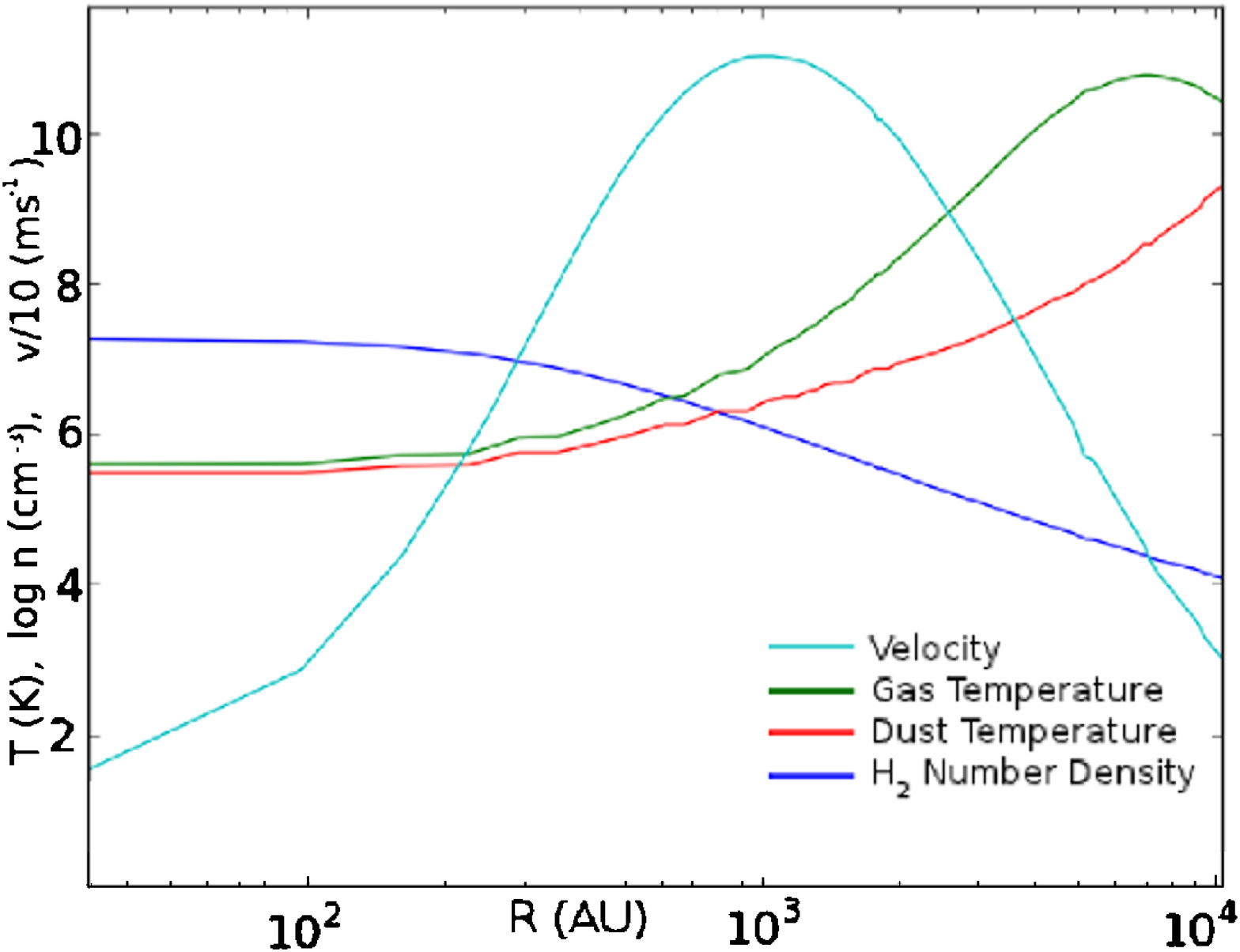}
 \caption{The spherically symmetric model of the pre-stellar core L1544 used as the envelope of the young protoplanetary disc in the hybrid model. Showing gas (green) and dust (red) temperature in kelvin, log number density (blue) in cm$^{-3}$ and inward velocity / 10 (cyan) in m$\,$s$^{-1}$. Adapted from KC2010 and Keto et al. (2013, in preparation).}
 \label{fig:l1544_model}
\end{figure}

To appropriately describe the environment within which a young protoplanetary disc is embedded, we combine the disc model from I2011 with a model of a dense core with characteristics similar to the well-studied pre-stellar core L1544 (\citealt{Keto2010}, hereafter KC2010). Although L1544 does not appear to contain a central protostar and disc itself, it is very similar in structure to L1521F,  another dense core in Taurus which hosts a low mass protostar and possibly a disc \citep{Bourke2006}. The KC2010 model follows the dynamical, chemical and thermal evolution of a contracting Bonnor-Ebert sphere \citep{Bonnor1956,Ebert1957} with total mass of 10\,M$_{\odot}$, until it reaches the density, temperature and velocity profiles which best match observations. The pre-stellar core model adopted here contains slight modifications compared to KC2010 due to the inclusion of oxygen cooling in the outer regions of the cloud (Keto et al. 2013, in preparation), where CO is mostly photodissociated \citep{Caselli2012}. Fig. \ref{fig:l1544_model} shows the physical parameters of the core model adopted here. \smallskip

\begin{figure}
 \includegraphics[width=84mm]{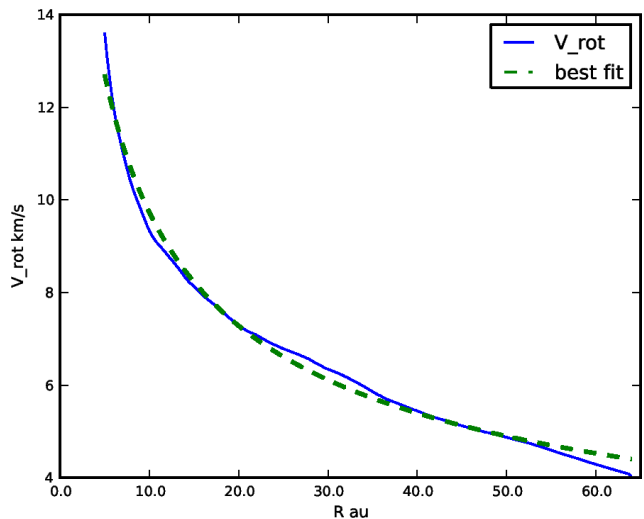}
 \caption{Azimuthally averaged rotational velocity in the disc mid-plane. The best fit curve follows the equation$V_{\rm rot}=\frac{31.5}{\sqrt{R+1.68}}+0.518$ .}
 \label{velocity}
\end{figure}

\begin{figure*}
 \includegraphics[width=168mm]{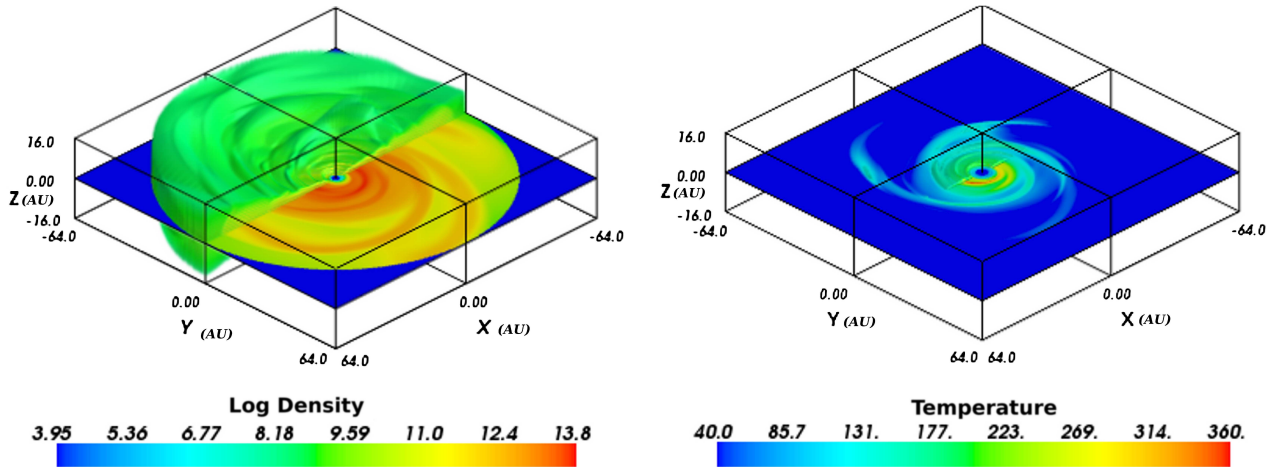}
 \caption{{\bf Left:} A 3D plot of log number density (cm$^{-3}$) showing the spiral structure in the xy plane and scale height of the disc. {\bf Right:} The 3D temperature (K) structure of the disc; regions cooler than 40$\,$K are not shown in 3D, in order to highlight the narrow central region containing hot material.}
 \label{rhoT} 
\end{figure*}

\begin{figure*}
 \includegraphics[width=168mm]{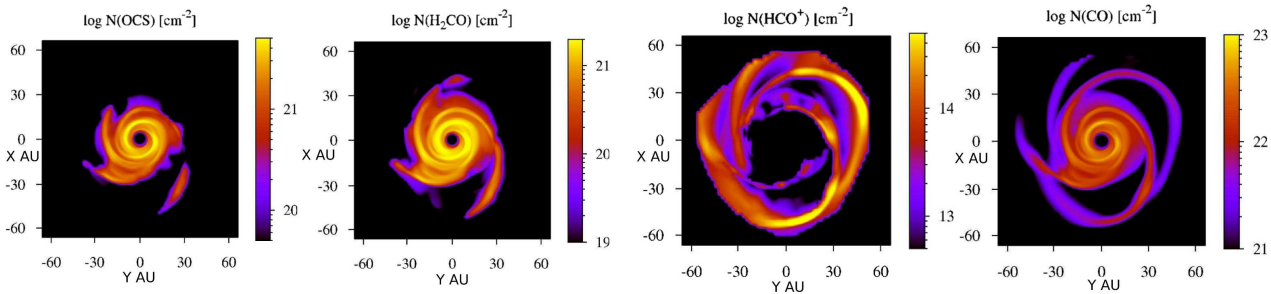}
 \caption{Column density distributions of OCS, H$_2$CO, HCO$^+$ and CO across the disc, as calculated by I2011. Note the different zones traced by the various species, with OCS mostly tracing the inner regions and HCO$^+$ probing the gas at larger radii.}
 \label{Chemistry} 
\end{figure*}

The pre-stellar core structure is maintained down to a radius of 80\,au, within which the model of the young protoplanetary disc is used. The disc structure is taken from the same hydrodynamics simulation of a 1$\,$M$_\odot$ protostar surrounded by a 0.39$\,$M$_\odot$ disc, described in I2011. Fig.\ref{velocity} shows the azimuthally averaged rotational velocity in the disc plane, together with the best fit curve, which will be used in Sec.~\ref{sec:alma_predictions} to compare with position velocity diagrams. The system is envisaged to be at a very early stage of evolution and embedded in an envelope.  The still growing protostar will eventually become an A star through accretion of most of the disc's mass. The hydrodynamics simulation is run with a proper equation of state for molecular hydrogen \citep{Boley2007}.  The protostar is allowed to move freely in response to disc torques, and radiative cooling is included \citep{Boley2009}. The self-gravitating disc exhibits prominent spiral structure, with H$_2$ number densities in the disc ranging from 10$^4$-10$^{13}$ cm$^{-3}$ and temperatures from 30-400 K (fig. \ref{rhoT}).  The disc model does not include magnetic fields, which may suppress gravitational instability and spiral wave formation by slowing down or even preventing early disc growth in the case of sufficient magnetisation and alignment between the parent cores rotation axis and magnetic field direction \citep{Joos2012,Li2013}. Thus, the observation of spiral structure would provide strong constraints on the dynamical state of discs. \smallskip 

For the non-LTE 3D radiative transfer modelling presented here, we interpolate the I2011 simulation output on to a 256$\times$256$\times$64 Cartesian grid with spatial resolution of 0.5$\,$au in x, y and z and sample from this to create the unstructured radiative transfer grid (see section \ref{subsec:gridding}).  As in the hydrodynamics simulation, the dust and gas temperatures are assumed to be in equilibrium within the disc and a gas to dust mass ratio of 100 is used throughout the entirety of the model. The dust opacities were adopted from \citet{Ossenkopf1994} and refer to dust grains with thick icy mantles and a 10$^6$ yr coagulation history.  Although dust grains within the protoplanetary disc are expected to coagulate further,  significantly affecting the opacity, we did not take grain growth into account in our model.  The dust opacities used in the radiative transfer are not the same as the ones used in the hydrodynamic model (which used the dust opacities of \citealt{DAlessio2001}). However, the results described here do not depend on the details of the dust opacity used during the hydrodynamic evolution.\smallskip

\subsection{Chemical structure} \label{subsec:chemical_structure}
Chemical abundances in the disc were taken from I2011, who followed gas-grain chemical processes during the dynamical evolution of the disc. Photochemistry is not taken into account, based on the assumption that in the early (Class 0, early ClassI) stages of protostellar accretion and evolution considered here, the protostar/disc system is heavily embedded in a thick envelope of gas and dust. The combination of envelope infall onto the disc and powerful outflows (see e.g. \citealt{Machida2013}) may drastically limit the illumination of the upper layers of young discs by the central protostar. Observations are needed to shed light on the importance of photochemistry at these early stages.

As described in I2011, the hydrodynamical simulation included Lagrangian tracer fluid elements in addition to the solution of the equations of hydrodynamics on an Eulerian grid.  These fluid elements were used to record the thermal history of the gas as material passed through and between the spiral structure. I2011 used this time evolution information from fluid elements to calculate the abundances of 125 species, related by 1334 reactions.  These abundances were then interpolated by I2011 on to a Cartesian grid with cell sizes 2.2$\times$2.2$\times$0.22$\,$au$^3$ in x, y and z to reveal the morphology of the disc as seen by different molecular species. From these we selected the four species which appear to trace different regions of the disc: the inner 30\,au (OCS), the inner 40\,au (H$_2$CO),  the region between $\simeq$40 and 60\,au (HCO$^+$) and the entirety of the disc (C$^{17}$O) (see fig. \ref{Chemistry}). The abundance model used for C$^{17}$O was the CO model reduced by a factor of 1792, the ratio of $^{16}$O to $^{17}$O in the local ISM \citep{Wilson1994}. As explained by I2011,  H$_2$CO and OCS mostly probe the central warm regions, where icy mantles evaporate, whereas HCO$^+$ preferentially traces the outer spiral pattern as in the central region it is destroyed by water molecules and transformed into H$_3$O$^+$ and CO. The simple chemistry in the KC2010 model, adopted here as the envelope of the protoplanetary disc, does not provide detailed abundances of molecular species (besides CO and H$_2$O, see also \citealt{Caselli2012}). As discussed in the results section \ref{sec:model_results}, abundances in the envelope have been adopted based on values measured toward similar objects. The molecular data used here for the radiative transfer are taken from the Leiden Atomic and Molecular DAtabase (LAMDA) (\citealt{Schoier2005} http://home.strw.leidenuniv.nl/$\sim$moldata/). \smallskip

\subsection{The radiative transfer code} \label{subsec:radiative_transfer_code}
\begin{figure}
 \includegraphics[width=84mm]{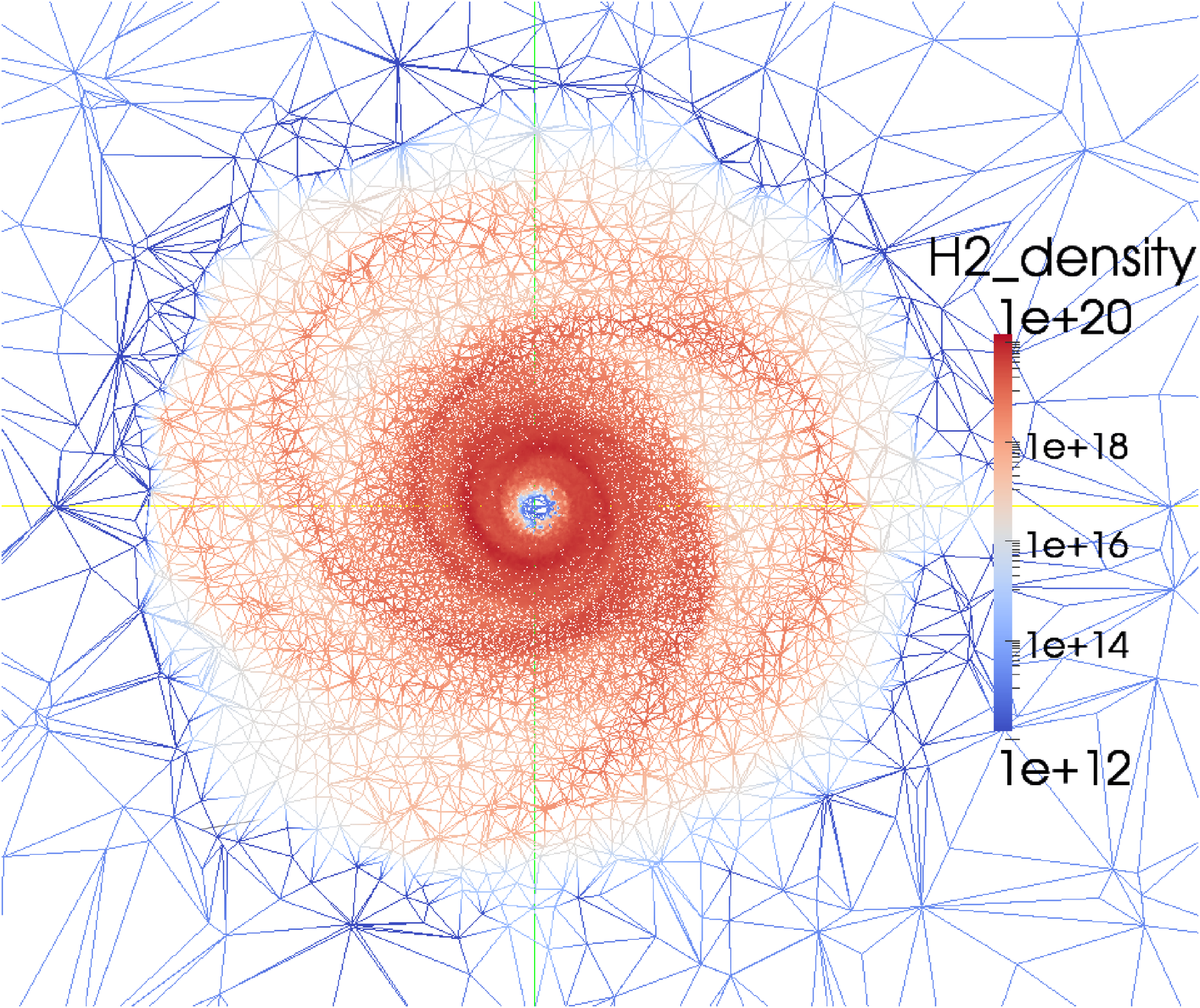}
 \caption{A plot of the points selected by the gridding process and the paths down which photons can propagate for points in the central r, $\theta$ plane. The points are colour coded by the density distribution (in m$^{-3}$, as used in LIME) and are more concentrated at small radii and in the most dense regions. The circular high density region in the centre is the disc model and is 128$\,$au in diameter.}
 \label{grid}
\end{figure}

\begin{figure}
 \includegraphics[width=84mm]{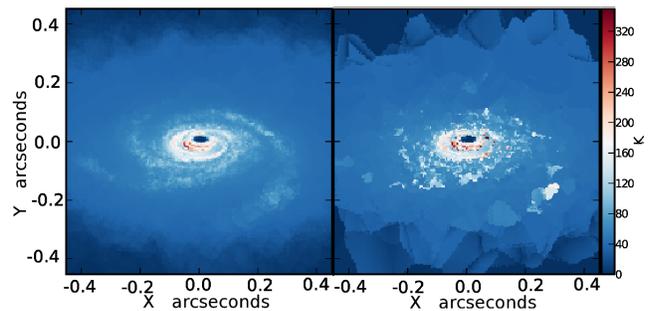}
 \caption{{\bf Left:} A 300$\,$GHz continuum image of the model, created from the average of ten LIME runs. {\bf Right:} the output of a single radiative transfer simulation, with artefacts due to finite gridding.}
 \label{averages}
\end{figure}

The radiative transfer program used is LIME (LIne Modelling Engine; \citealt{Brinch2010}), which  calculates line intensities based on a weighted sample of randomly chosen points in a continuous 3D model. The method of selecting these points is given in section \ref{subsec:gridding}. At each of these points, the density of the main collision partner (H$_2$), gas and dust temperatures, velocity, molecular abundances and unresolved turbulent velocity are specified. These points are then smoothed by Lloyd's algorithm \citep{Lloyd1982} in order to minimise the variation in distance between points whilst keeping the same underlying distribution. These points are then connected by Delaunay triangulation and it is between the points connected by this method that photons are allowed to propagate (fig. \ref{grid}). The level populations of the selected molecules are calculated at each of these points from collisional and radiative (de)excitation and the local radiation field is calculated. This is repeated 20 times with the populations of each level converging towards a single value. This number of iterations is sufficient for the signal to noise ratio of the level populations (as defined in \citealt{Brinch2010}) to exceed 1000 for 99\% of the points, ensuring that the simulation has converged on a stable level population. After 20 iterations the model is ray-traced in order to produce synthetic brightness maps. The average of ten separate runs was taken to minimise the artefacts in the output images, resulting from the grid construction (Fig. \ref{averages}).

\subsection{Grid construction} \label{subsec:gridding}
 In order to construct the grid, candidate points are randomly selected from the volume to be simulated. These candidate points are associated with values of the H$_2$ number density and the number density of the molecular species being simulated at the position of the point in the underlying model. These are then compared against the H$_2$ and molecular density of a reference point in order to decide if they will be included in the grid. 

\begin{figure}
 \includegraphics[width=84mm]{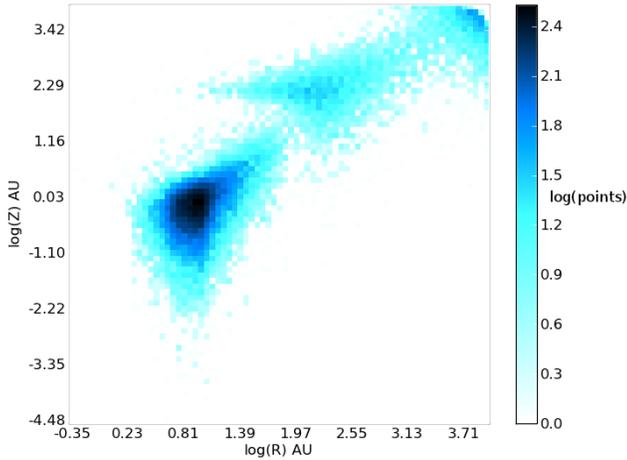}
 \caption{A 2D histogram of the point distribution throughout the model. The disc and envelope can be seen as two separate entities which have to be sampled using different point distributions.}
 \label{points}
\end{figure}

Candidate grid points are selected at random in a cylindrical coordinate system that is linearly spaced in z and $\theta$ and logarithmically spaced in r. For each point to be selected, a random number $\alpha$ is drawn from the semi-open set [0,$\,$1) to be used as a threshold. After selection of random coordinates, the hydrogen density and molecular density at the candidate point (n and m, respectively) are compared against the densities of a reference point on the inner edge of the disc (n$_0$ and m$_0$). If $\alpha<\left( \frac{n}{n_0} \right)^{0.3}$ or $\alpha< \left( \frac{m}{m_0} \right)^{0.3}$ then the point is selected for use. Otherwise another r, $\theta$, z co-ordinate is selected and it becomes the candidate point. 20\% of these selected points are forced to be at radii greater than $\sqrt{R_{min}R_{max}}$ (where $R_{min}$ and $R_{max}$ are the inner and outer radii of the model) in order to stop too many of the selected points clustering in the high density disc and leaving the envelope under sampled. In addition to this method of selection, 5\% of the points are linearly distributed in x, y and z with no bias with regards to density or abundance. This provides a minimum level of sampling for the large low density regions in the outer parts of the simulated volume. See fig. \ref{points} for an example of the points distribution in r, z. The function comparing the candidate point to the reference point and the candidate point distribution were selected empirically to sample all scales while ensuring that the majority of points are located in the inner disc where the density is higher.  \smallskip

\section{Model Results} \label{sec:model_results}

Simulations are limited to molecules for which we have abundances in both the disc and envelope, and also have reliable Einstein and collisional coefficients. In the envelope, the HCO$^+$ abundance profile follows the H$_2$O profile of L1544 \citep{Caselli2012}, scaled so the maximum is 1$\times$10$^{-8}$. For H$_2$CO, a step profile was considered, with a fractional abundance of 1.5$\times$10$^{-8}$ at radii greater than 0.04$\,$pc and 1.5$\times$10$^{-9}$ at smaller radii \citep{Young2004}. For OCS we adopted a constant abundance of 1.9$\times$10$^{-9}$ \citep{Ren2011}. The C$^{17}$O abundance profile in the envelope follows the CO profile calculated for L1544 by KC2010, scaled by the $^{17}$O/$^{16}$O isotope abundance. These estimates for the abundances in the envelope are simplistic. However, as seen in section \ref{sec:alma_predictions}, the envelope contribution to the line is very limited in velocity and can be spectrally disentangled from the disc contribution.\smallskip

We focus on the frequency range available with ALMA, with particular attention to band 7, which offers the best trade off between resolution and sensitivity. The results presented in this section are limited to those lines with detectable emission/absorption which can be used to trace either spiral structure or rotation. Thus, we consider C$^{17}$O(3$\rightarrow$2), HCO$^+$(3$\rightarrow$2), OCS (26$\rightarrow$25) and H$_2$CO(4$_{04}\rightarrow$3$_{03}$).  The frequencies and upper level energies of these transitions are given in table \ref{sigmas}. \smallskip

\begin{figure*}
 \includegraphics[width=150mm]{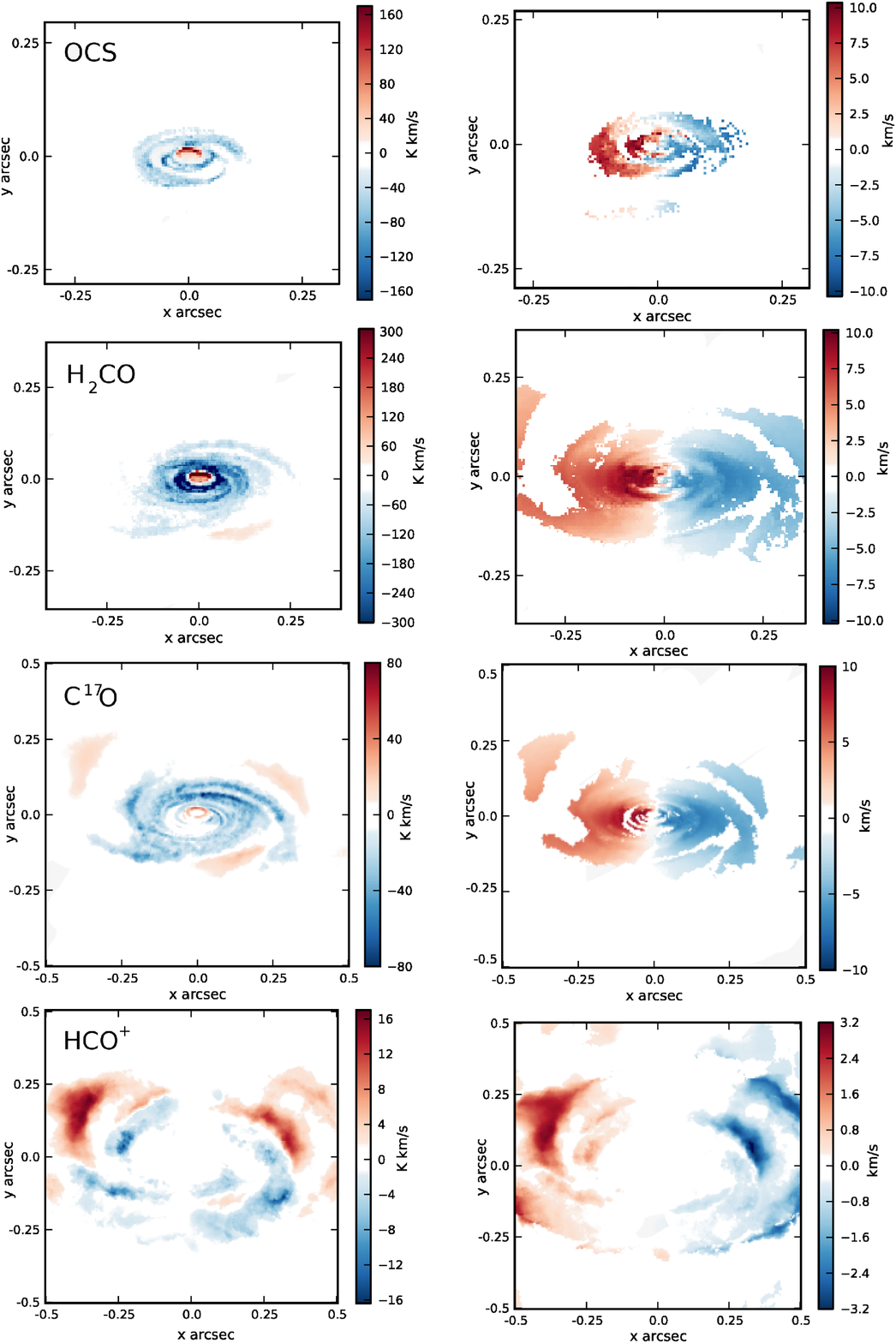}
 \caption{{\bf Left:} Continuum subtracted integrated intensity maps. {\bf Right:} Intensity weighted velocity maps. The spiral structures in different regions of the disc are highlighted by different molecular lines. From top to bottom the lines displayed are: OCS 26$\rightarrow$25, H$_2$CO 4$_{04}$$\rightarrow$3$_{03}$, C$^{17}$O 3$\rightarrow$2 and HCO$^+$ 3$\rightarrow$2. All maps are obtained from integration over -12.5 to -0.5 and 0.5 to 12.5 km$\,$s$^{-1}$ in order to avoid the envelope contribution.}
 \label{sim_all}
\end{figure*}

To simulate observations, the model was placed at roughly the distance of nearby low-mass star forming regions (100$\,$pc). The inclination angle was varied from 15$^\circ$ to 75$^\circ$ relative to the edge on case. From these simulated observations, integrated intensity maps, intensity weighted velocity maps and position velocity diagrams were created. The integrated intensity and  intensity weighted velocity maps (figs. \ref{sim_all}, and \ref{mom0_maps}) are integrated from -12.5 to -0.5$\,$km$\,$s$^{-1}$ and +0.5 to +12.5$\,$km$\,$s$^{-1}$ to avoid domination by the contribution from the envelope, which can be seen in some PV diagrams as the strong absorption feature at all positions around zero velocity (see section \ref{sec:alma_predictions}). Intensity weighted velocity maps are shown with a cut-off of 3$\sigma$ as described in section \ref{sec:alma_predictions}.\smallskip

OCS 26$\rightarrow$25 traces only the innermost 20$\,$au of the disc and can be used to examine the central regions (fig. \ref{sim_all}). As OCS is not seen in outflows (e. g. \citealt{Stanke2007}, \citealt{VDTak2003}), it can be used to measure the rotation of the central part of the disc without contamination from shocked material along the outflow. The OCS lines are unique amongst the lines simulated in that they trace radii $<$16$\,$au, even smaller than the extent of the OCS column density distribution shown in fig. \ref{Chemistry}. This is because the OCS line, which has the highest upper energy level among the selected transitions, traces only the hottest and densest regions of the disc, where the high energy states are populated. As a result, the innermost spiral structure is best seen in OCS (26$\rightarrow$25), as shown in fig. \ref{sim_all}. The OCS line is mostly seen in absorption against the bright continuum emission from the disc mid-plane. The exception to this is towards the centre of the disc, where the hole in the hot, dense mid-plane produces little continuum to be absorbed. However, this may be an artefact of our treatment of the inner hole, given that we neglect the presence of the protostar and other hot material within the central 2$\,$au. Thus, we expect absorption also to be observed toward the central region of the disc.\smallskip

The abundance of the H$_2$CO molecule traces the spiral structure of the inner $\sim$40$\,$au of the disc, which can be seen in the integrated intensity map, and its rotation is detected out to larger radii compared to the OCS line (fig. \ref{sim_all}). As with the OCS line, the H$_2$CO 4$_{04}$$\rightarrow$3$_{03}$ line is mostly seen in absorption against the disc mid-plane continuum, with the central region seen in emission. However, unlike the OCS, the H$_2$CO extends out to large enough radii to trace the voids between the outer spiral arms. In these regions, H$_2$CO line emission can be seen.\smallskip

The fractional abundance of C$^{17}$O is constant at 2$\times$10$^{-8}$ across the disc, so that the selected C$^{17}$O line is the most accurate in reproducing the physical structure of the disc. Like H$_2$CO, C$^{17}$O shows emission in voids between spiral arms. The C$^{17}$O line is visible from the inner edge of the disc to its outer edge. Like OCS, C$^{17}$O is not seen in outflows \citep{Yildiz2012}, so we do not expect contamination.\smallskip

The HCO$^+$ line only traces the outer regions of the disc. In fig. \ref{sim_all} we can see that the HCO$^+$ line shows the outer edges of the spiral arms in absorption, but it also shows emission from the diffuse gas further out in the disc. This allows the velocity structure of the disc to be measured out to larger radii, allowing constraints to be placed on the rotation of the disc out to larger radii than with other molecules. 

\begin{figure*}
 \includegraphics[width=110mm]{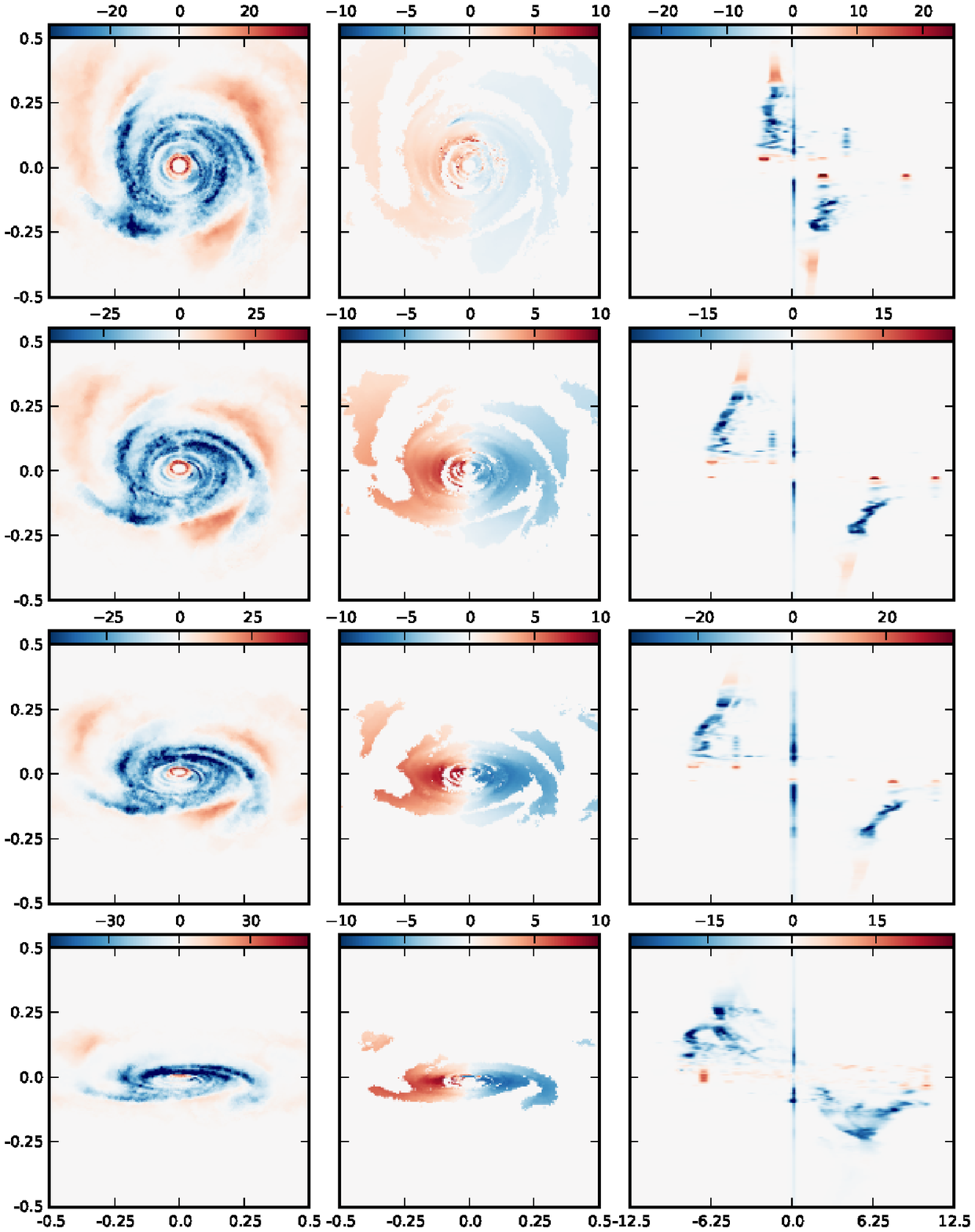}
 \caption{C$^{17}$O 3$\rightarrow$2 {\bf left:} Continuum subtracted integrated intensity map (in K$\,$km$\,$s$^{-1}$). {\bf Centre:} Intensity weighted velocity map (in km$\,$s$^{-1}$). {\bf Right:} Position-velocity diagram with intensities in K. The inclinations for which results are shown from top to bottom are: 75$^\circ$, 45$^\circ$, 30$^\circ$ and 15$^\circ$ to edge on.}
 \label{other_inc}
\end{figure*}

Simulations at inclinations other than 30$^\circ$ were also performed with C$^{17}$O. Fig. \ref{other_inc} shows the C$^{17}$O 3$\rightarrow$2 line in the same physical model but at 15$^\circ$, 30$^\circ$, 45$^\circ$ and 75$^\circ$ to edge on. At 15$^\circ$ to edge on, some structure in the integrated intensity map is still observable and the spiral pattern is clearly seen in the position velocity diagram. At larger inclinations it gets more difficult to obtain any kinematic information but the physical structure of the disc becomes clearer.

\section{Predictions for ALMA} \label{sec:alma_predictions}

In order to make predictions of the observability of the features for which we have constructed synthetic brightness maps, the Common Astronomy Software Applications package (CASA) was used to simulate observations with the completed ALMA in the 26th most extended configuration (out of a total of 28). The longest baseline of this configuration is 14.4\,km and the beam sizes range between 0.02 and 0.03 arcseconds at the selected frequencies. This angular resolution is close to the size of the spiral structure in the disc. As the weakest lines studied in the model are of the order of 0.1-0.2$\,$mJy$\,$beam$^{-1}$ at this resolution, Band 7 sensitivities of the order of 0.02$\,$mJy$\,$beam$^{-1}$ are required. This implies around 6 hours of integration time. The exact sensitivities are given in Table \ref{sigmas}.
\begin{table*}
  \centering
  \begin{minipage}{90mm}
    \caption{Line sensitivities obtained with ALMA after 6 hours integration time with velocity resolution of 400$\,$m\,s$^{-1}$, obtained with the ALMA on-line sensitivity calculator.}
    \label{sigmas}
    \begin{tabular}{c||c|c|c|c}
      \hline
      Species & Transition & Frequency & Upper energy level & Sensitivity\\
              &            &   (GHz)   &       (K)          & (mJy beam$^{-1}$) \\
      \hline
      C$^{17}$O & 3$\rightarrow$2 & 337.06 & 32.3 & 0.0270 \\
      OCS & 26$\rightarrow$25 & 316.15 & 205 & 0.0275 \\
      H$_2$CO & 4$_{04}\rightarrow$3$_{03}$ &  290.62 & 34.9 & 0.0228 \\
      HCO$^+$ & 3$\rightarrow$2 & 267.56 & 25.7 & 0.0176 \\
      \hline
    \end{tabular}
  \end{minipage}
\end{table*}
In order to clean the images which featured mainly absorption, the sky model used was the output of the LIME simulations with the continuum subtracted and then the spectrum inverted. The simulated images were cleaned down to a 3$\sigma$ level with $\sigma$ (the sensitivity) given in Table \ref{sigmas}.\smallskip

\begin{figure}
 \includegraphics[width=84mm]{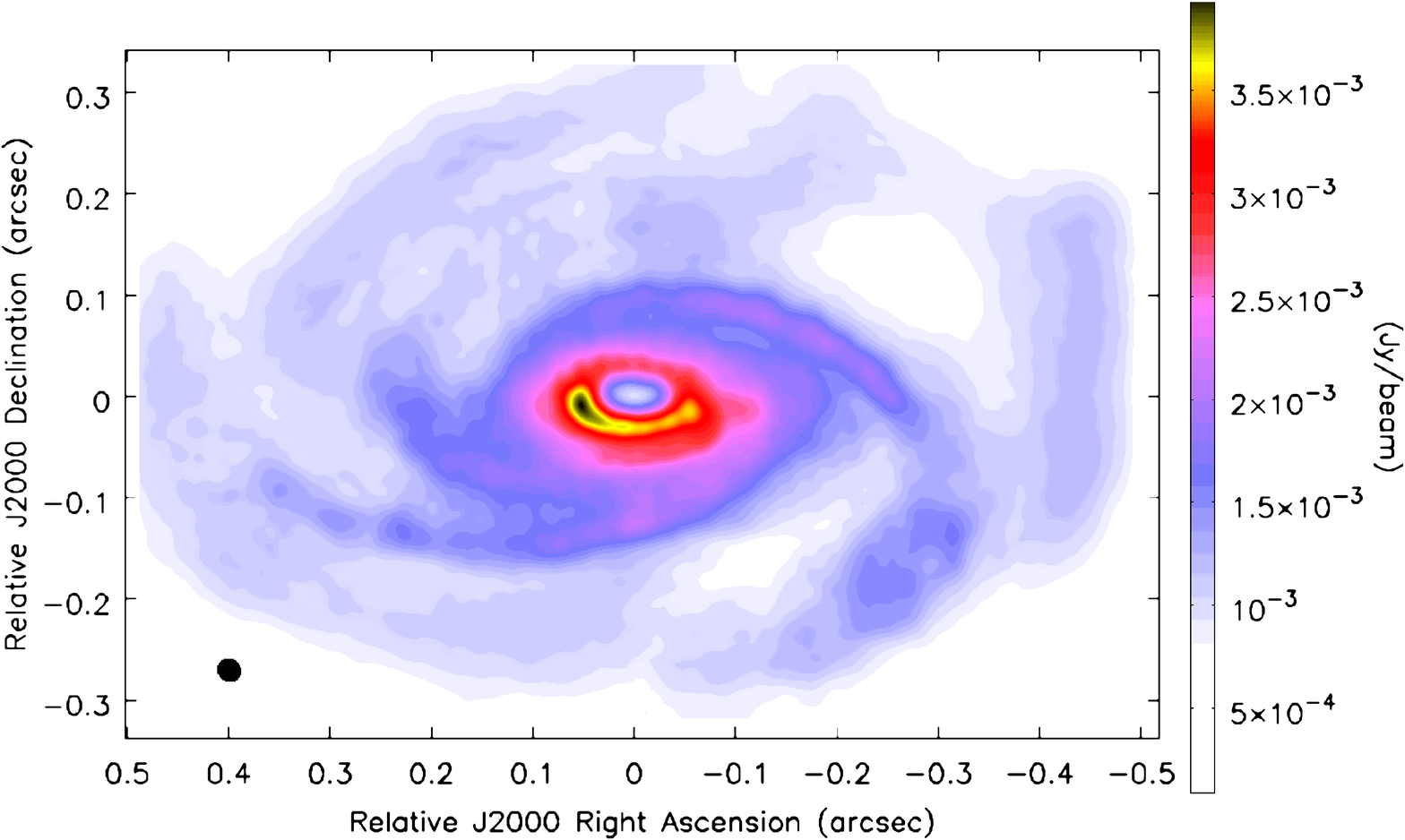}

 \caption{CASA simulation of the continuum emission of the model at 300GHz, using the same ALMA configuration as used for the molecular lines. The size of the beam at 300GHz is shown in the lower left corner.}
 \label{continuum}
\end{figure}

\begin{figure*}
 \includegraphics[width=168mm]{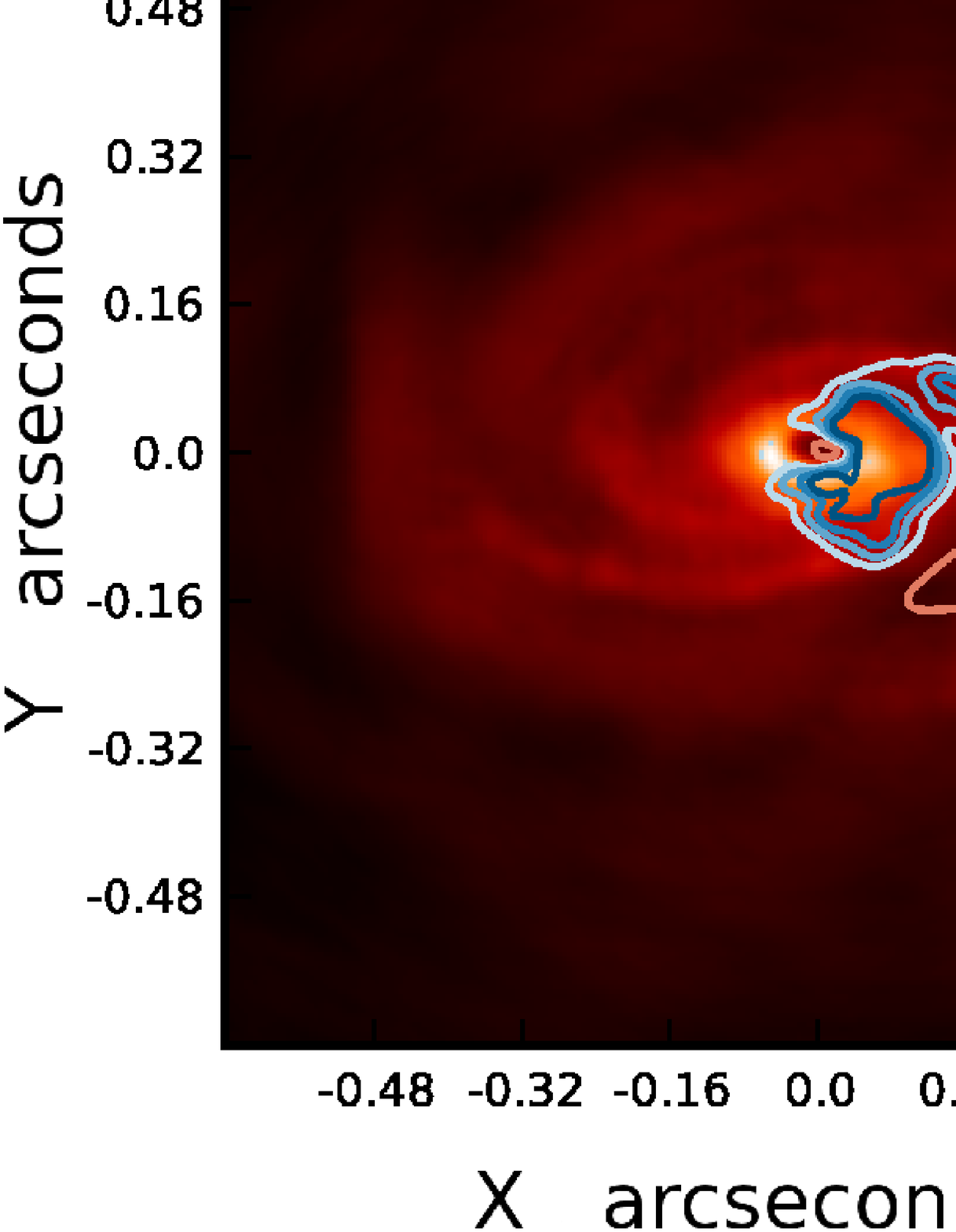} 
 \caption{H$_2$CO 4$_{04}\rightarrow$3$_{03}$ channel maps as simulated with CASA. The contours start at 0.35 mJy$\,$beam$^{-1}\,$km$\,$s$^{-1}$  and are in steps of 0.35 mJy$\,$beam$^{-1}\,$km$\,$s$^{-1}$ (absorption in blue, emission in red). Overlaid on the 1mm continuum emission. The integration range for each map is 2.4 km$\,$s$^{-1}$, for the top left the range is from -10.8 to -8.4 km$\,$s$^{-1}$ and for the bottom right the range is from 8.4 to 10.8 km$\,$s$^{-1}$.}
 \label{h2co_chanmap}
\end{figure*}

Fig. \ref{continuum} shows the 1\,mm continuum emission as observed with the chosen ALMA configuration. The spiral structure of the disc is clearly visible at this inclination and and it becomes clearer at angles closer to face-on. A mass estimate for the disc was calculated using the equation $M=g S_\nu d^2/\kappa_\nu B_\nu(T_d)$ where $g$ is the gas/dust mass ratio, $S_\nu$ is the integrated flux, $d$ is the distance to the source, $\kappa_\nu$ is the dust opacity and $B_\nu(T_d)$ is the black body function for a given dust temperature \citep{Beltran2006}. Using the above equation and integrating the flux from y=0.2 to -0.2 and from x=0.4 to -0.4, we obtain a disc mass of 0.1 and 0.005$\,$M$_\odot$, if the dust temperature is fixed at 40 and 400\,K, respectively. This shows that the disc is optically thick to millimetre wavelengths and millimetre observations underestimate the disc mass significantly, by factors of 4 or more (see also \citealt{Forgan2013}). In young, massive discs such as those we are investigating, longer wavelength continuum observations are necessary for determining the disc mass.\smallskip

Fig. \ref{h2co_chanmap} shows a channel map of the H$_2$CO 4$_{04}\rightarrow$3$_{03}$ line overlaid on the 1$\,$mm continuum emission. From this it can be seen that the absorption features trace the disc spiral structure whilst the line emission emanates from gaps in the outer disc, where there is less continuum to be absorbed. The central panel is dominated by the envelope contribution, causing strong absorption throughout the disc. The other panels show the line following a rotation pattern distorted by the presence of the spiral arms. \smallskip

\begin{figure}
 \includegraphics[width=84mm]{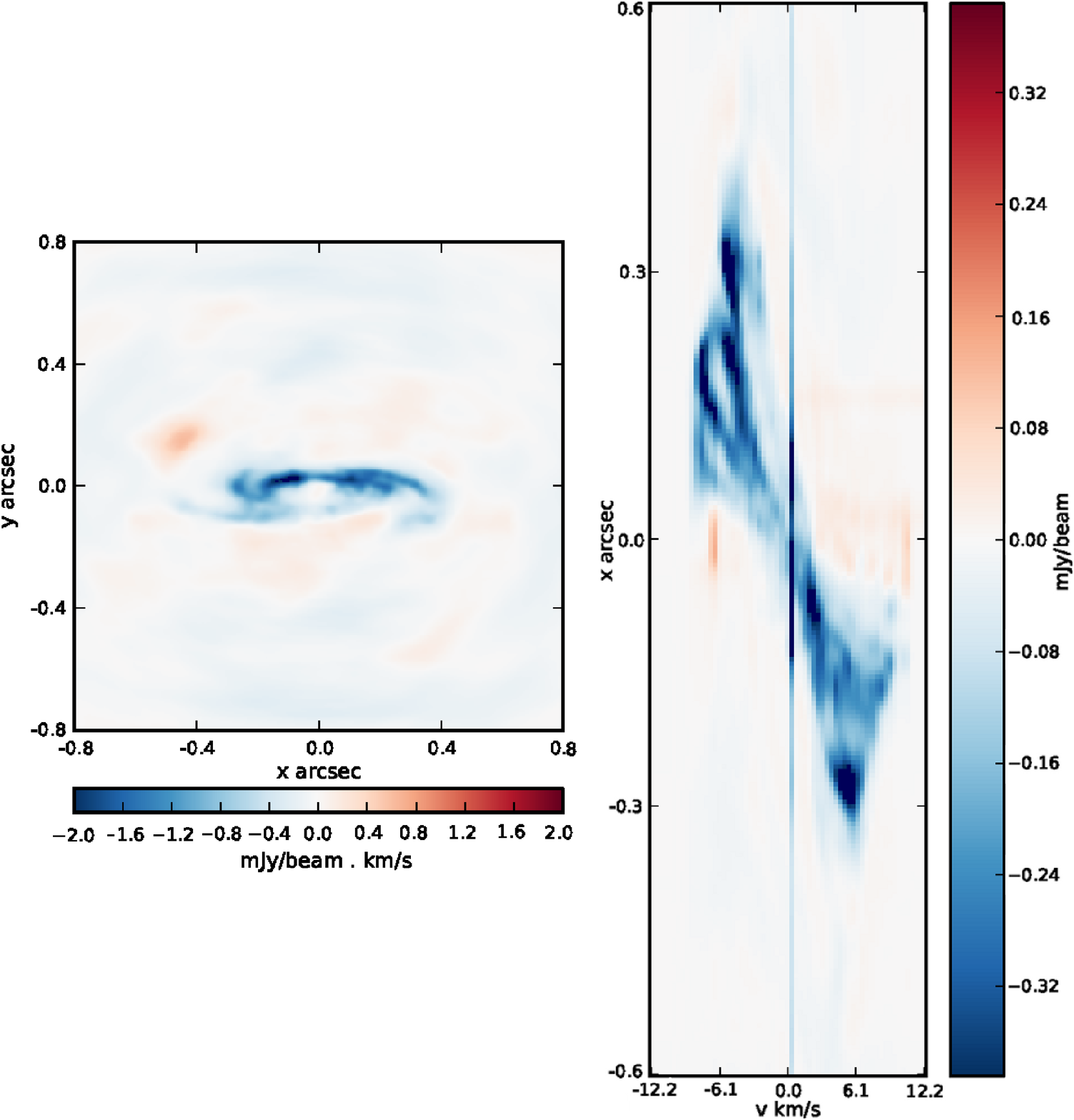}

 \caption{CASA simulation of the C$^{17}$O 3$\rightarrow$2 line with the disc inclined at 15$^\circ$ to edge on. {\bf Left:} Integrated intensity map. {\bf Right:} Position velocity diagram, where the spiral structures can be seen as narrow finger-like structures extending from the origin.}
 \label{15deg}
\end{figure}

Fig. \ref{15deg} shows the CASA simulation of the C$^{17}$O 3$\rightarrow$2 integrated intensity and PV diagram for an inclination angle of 15$^\circ$ to edge on.  Spiral structure can be clearly seen in the position velocity diagram even at low inclinations. The ``finger-like'' structures in the PV diagram are similar to those seen in observations of the Milky Way and in simulations of spirals in the Milky Way (e. g. \citealt{Bissantz2003}). Although the detail of the spiral structure (number of arms, pitch angle etc.) is difficult to gauge from such a PV diagram, high spectral and angular resolution observations can definitely unveil their presence and thus be used to find gravitationally unstable discs. \smallskip

\begin{figure*}
 \includegraphics[width=140mm]{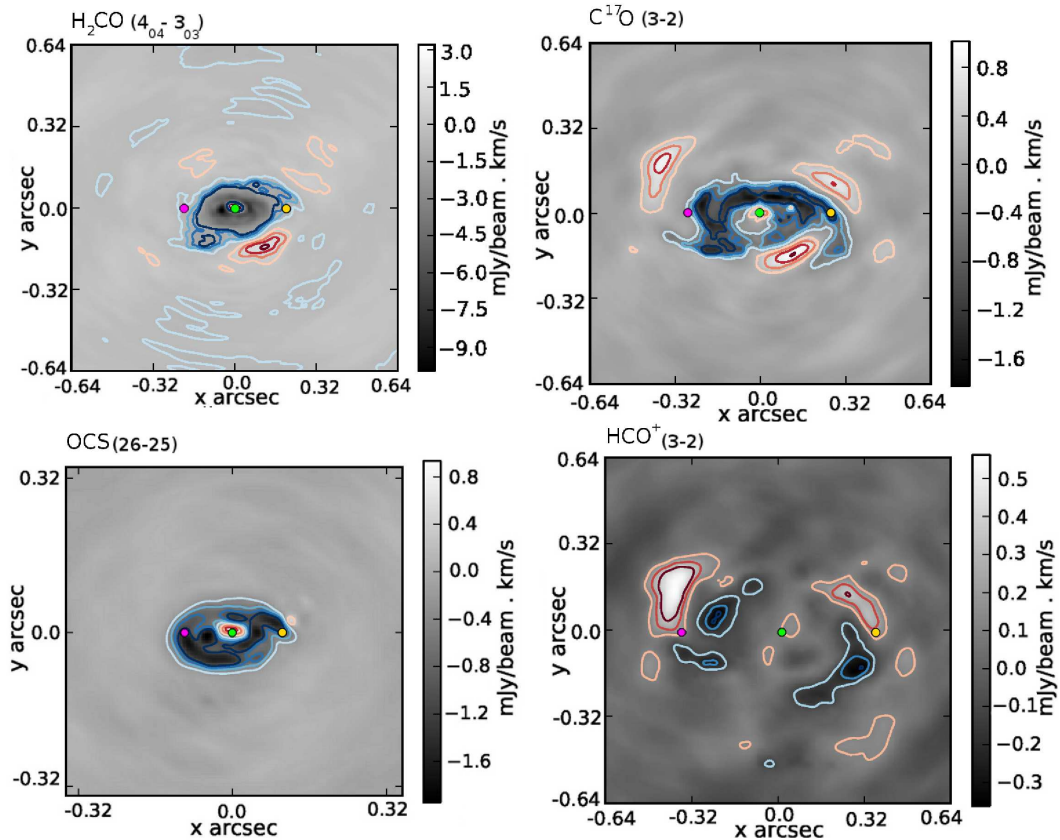}
 \caption{Integrated intensity maps for the simulated observations of: {\bf Top Left:} H$_2$CO 4$_{04}\rightarrow$3$_{03}$, {\bf Top Right:} C$^{17}$O 3$\rightarrow$2, {\bf Bottom Left:} OCS 26$\rightarrow$25 and {\bf Bottom Right:} HCO$^+$ 3$\rightarrow$2. Contours start at 3$\sigma$ and increase in intervals of 3$\sigma$ for each of the maps except the H$_2$CO contours, which start at the 5$\sigma$ level and increase in intervals of 5$\sigma$. Emission is in red contours, absorption in blue. The coloured dots refer to the positions from which the spectra shown in fig. \ref{spectra} are taken.}
\label{mom0_maps}
\end{figure*}

Fig. \ref{mom0_maps} shows the simulated integrated intensity maps of the disc in the four selected molecular transitions. As expected from the radiative transfer results, observations of these transitions can probe different regions of the disc, from the OCS line tracing the innermost regions, to the HCO$^+$ line showing the outer regions. Thus, the high sensitivity and angular resolution of ALMA will allow detection of spiral structure both in the continuum emission as well as  in appropriately selected molecular transitions.\smallskip

\begin{figure*}
 \includegraphics[width=180mm]{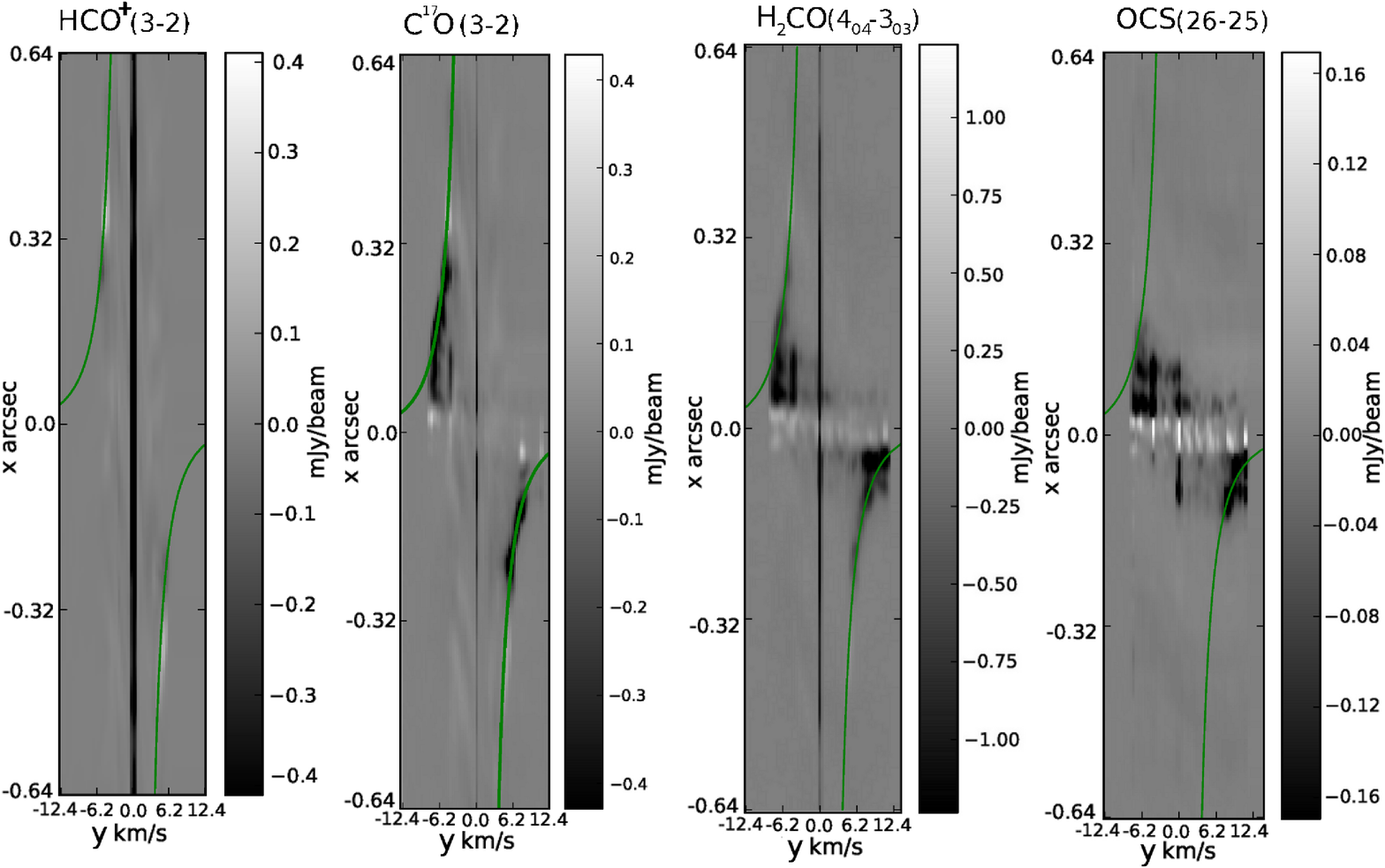}
 \caption{Position velocity diagrams for each of the lines simulated with CASA. The green curve on each diagram is the best fit curve to the azimuthally averaged rotation profile from fig. 4.}
 \label{pvs}
\end{figure*}

Position-velocity diagrams are shown in fig. \ref{pvs}. These diagrams can be used to reconstruct the rotation of the disc. The green curves in each panel of fig. \ref{pvs} display the best fit rotation curve for the physical model (see fig. \ref{velocity}), showing that the rotation profile of the disc can be constrained by kinematic information gathered from species tracing different regions of the disc.\smallskip

\begin{figure*}
 \includegraphics[width=168mm]{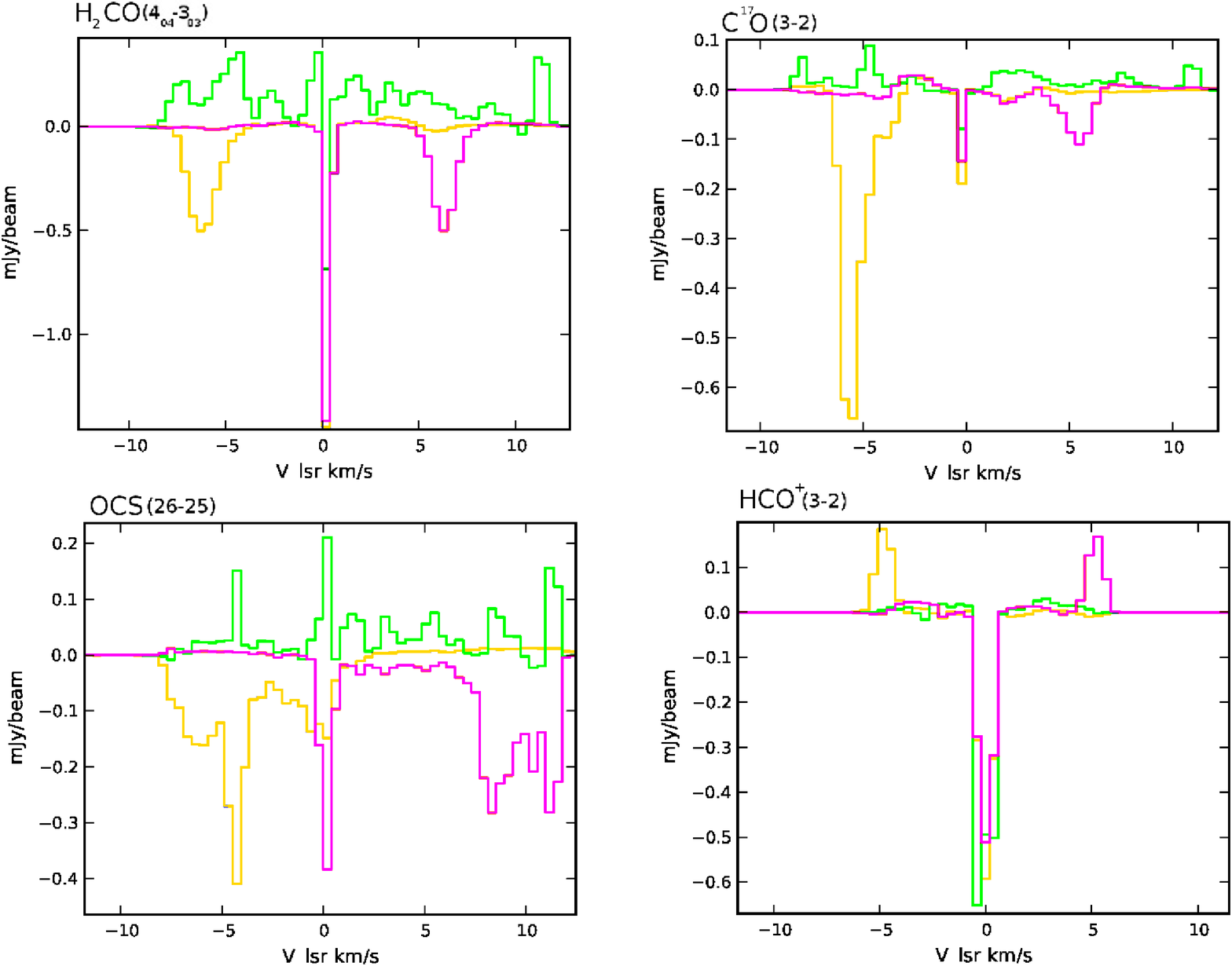}
 \caption{Simulated spectra for each of the selected transitions at the positions indicated by the coloured dots in fig. \ref{mom0_maps} (the pink, green and yellow curves are taken from the locations marked by the dots of the same colour). In each diagram, the three spectra are taken from three positions on the Y=0 axis, one is at X=0 (green) and the other 2 are from x=$\pm\,$N (plus yellow, minus pink) where N is the distance from the centre: 10 au for OCS, 20 au for H$_2$CO, 28 au for C$^{17}$O, 38 au for HCO$^+$.}
 \label{spectra}
\end{figure*}

The spectra presented in Fig. \ref{spectra} were extracted from the maps in fig. \ref{mom0_maps} for the positions indicated by the corresponding coloured dots along the Y=0 axis. As with all the CASA simulated images, these spectra were obtained with an ALMA window with total bandwidth 938$\,$MHz, giving velocity resolutions between 418 and 527\,m\,s$^{-1}$. Very notably, these spectral features are not symmetric about the disc centre. For example, C$^{17}$O(3--2) has a significantly stronger blueshifted absorption, due to the presence of a spiral arm along the same line of sight. In the OCS lines, the two sides are not moving with equal velocities and the line widths and shapes are markedly different. These can be attributed to the non-axisymmetric nature of the model.

\section{Conclusions} \label{sec:discussion}

In this paper we have presented radiative transfer simulations of a hybrid model comprising a 0.39$\, M_\odot$ self gravitating disc with a radius of 64$\,$au and spiral density waves, surrounded by an envelope that is a contracting 10$\,M_\odot$ BE-sphere. The main results of these simulations are as follows:\newline
$\bullet$ CASA simulations of this model show that at a distance of 100$\,$pc, extraction of kinematic and structural information from the continuum and molecular lines is possible with ALMA band 7.\\
$\bullet$ Our simulations show that molecular features are predominantly seen in absorption against the hot mid-plane of self-gravitating protoplanetary discs, with emission only being seen where voids in the structure of the disc do not provide a bright continuum source to be absorbed.\\
$\bullet$ The quiescent nature of the envelope around such discs only affects lines within $\pm\,$0.5 km$\,$s$^{-1}$ of the LSR velocity.\\
$\bullet$ The lines studied (OCS 26$\rightarrow$25, H$_2$CO 4$_{04}$$\rightarrow$3$_{03}$, C$^{17}$O 3$\rightarrow$2 and HCO$^+$ 3$\rightarrow$2) taken together allow all regions of the disc to be sampled and the rotation of the disc to be constrained from the inner edge to the outer edge.\\
$\bullet$ Spiral structures in young embedded discs are detectable at a wide range of inclination angles. They can be resolved spatially at angles close to face on and can be inferred from position velocity diagrams at low inclination angles.\\
 $\bullet$ The dust continuum emission at millimetre wavelengths is optically thick and millimetre observations significantly underestimate the disc mass, by a factor of 4 or more.\smallskip

One assumption underpinning this model is that the gas and dust are in thermal equilibrium in the disc. If they are not and the dust is significantly cooler than the gas then transitions may not show up in absorption. However, the large volume densities are expected to provide efficient dust-gas coupling. Outflows could contaminate line profiles of  those species  which commonly trace outflows, such as CO and HCO$^+$. In these cases, a detailed study of their line profiles using high spectral resolution should help to disentangle the various kinematic components.  In the future we will study this, including outflow lobes in the disc/envelope system.  However species such as OCS and C$^{17}$O, which are not commonly seen in outflows, can be used to trace disc rotation and structure.\smallskip

These results demonstrate that ALMA will be able to probe the very earliest stages of circumstellar discs which feed the growing protostar and eventually will evolve in a planetary system. Our results show that ALMA observations in band 7 will allow the assessment of whether or not these young discs are gravitationally unstable and test theories of pre-stellar to protostellar core evolution. If such gravitational instabilities exist, they will have significant implications for the evolution of the disc. For example, they will affect fragmentation, planetesimal formation \citep{Boley2009,Johnson2013,Gibbons2012} and the chemical evolution in the disc (I2011). 

\smallskip

\section*{Acknowledgements}
PC and TWH acknowledge support of successive rolling grants awarded by the UK Science and Technology Funding Council.  
TAD and JDI acknowledge studentships from the Science and Technology Facilities Council of the United Kingdom (STFC), JDI also acknowledges funding from the European Union FP7-2011 under grant agreement no 284405. 
TAD would like to thank Nichol Cunningham for her assistance with CASA and Christian Brinch for his radiative transfer code LIME and for guidance on using it.
We also thank the anonymous referee whose comments improved the initial draft of this manuscript.

\bibliographystyle{mn2e} 
\bibliography{new_bib}
\bsp
\label{lastpage}
\end{document}